\documentstyle[12pt]{article}
\begin{document}
\baselineskip=18pt
\def\be{\begin{equation}}
\def\ee{\end{equation}}
\def\E{{\rm e}}
\def\e'{e^\prime}
\def\p{^\prime}
\vskip 1cm
\begin{tabbing}
\hskip 11.5 cm \= CERN-TH/96-329\\
\>November 1996
\end{tabbing}
\vskip 1cm
\begin{center}
{\Large\bf Operator solutions of two-dimensional chiral gauge theories}
\vskip 1.1cm
{\large \bf E. Abdalla}$^{a,b}$\footnote{Permanent address: Instituto de 
F\'\i sica-USP, C.P. 66.318, S. Paulo, Brazil.} \\
\vskip 0.3cm
{{\it $^a$CERN, Theory Division}\\
{\it $^b$ICTP, High-Energy Division}\\
\vskip 0.3cm
$^b$ elcio@ictp.trieste.it}
\end{center}
\abstract 

\noindent
The exact operator solutions of two-dimensional anomaly-free 
chiral abelian gauge theories are obtained. We show that 
anomaly-cancellation conditions arise as consistency 
requirements of these solutions. For a certain class of 
flavour symmetries, fermion condensates are constructed.
These are shown to violate the fermion-number conservation rule. 
The models are extended to include massive fermions. We propose 
a bosonised lagrangian for the massive theory and verify that
it complies with the Gupta-Bleuler condition.

\vskip .3cm
\noindent
CERN-TH/96-329\\
November 1996

\vfill\eject

\section{Introduction}
\indent

Two-dimensional gauge theories have provided us with valuable
informations since the pioneering work of Schwinger on two-dimensional
quantum electrodynamics \cite{sch}. 
The subject evolved rapidly with the discovery of the operator solution 
of this model \cite{ls}. 
It was subsequently realised that this solution is of 
a key importance for the understanding 
of fundamental physical problems \cite{ls,cks} such as confinement 
and the structure of the $\theta$-vacuum. 

Any operator solution of this kind can only be of 
physical relevance if it respects gauge symmetries.
In chiral gauge theories, the gauge field couples to a 
non-conserved current and the gauge symmetry is broken.
In Salam-Weinberg model, different multiplets of
right and left chiralities are introduced in order to restore 
the gauge symmetry. The gauge anomalies only cancel for a certain 
combination of multiplets. This anomaly-cancellation condition
lead to the discovery of the top quark. 

In this paper, we obtain the operator solution of a general 
two-dimensional quantum theory with
$SU_R(n)\times SU_L(q)$ symmetry. As in Salam-Weinberg 
model, the anomalies cancel between different multiplets. 
We start by deriving the anomaly-cancellation
condition in Section 2. 
In Section 3, we write the gauge and the
fermion fields in terms of bosonic variables. 
The full operator solutions are then found by 
requiring that these fields obey
Dirac and Maxwell equations. 
These solutions are similar to the operator solutions of the 
flavoured Schwinger model \cite{aar}. Massive excitations are
obtained from the massive mesons. These are, in turn, 
constructed from the gauge field and their masses are generated by
a dynamical Higgs mechanism.
The massless operators build the $\theta$ vacua which are labelled by
two angles $\theta_1$ and $\theta_2$. 
Out of these solutions, condensates with 
non-vanishing fermion numbers are also constructed.
We calculate the two-point 
functions of these condensates. The non-vanishing result shows that 
the fermion-number conservation rule is violated. 
In Section 4, we study the massive theory.
We propose an ansatz for the bosonised massive lagrangian, using the principle
of form invariance. This lagrangian depends on free parameters. In order to
fix them, we identify the combinations of the fields that are required to 
remain massless by the BRST condition. 
This then insures the validity of our lagrangian.
The last section is devoted to conclusions and comments.

\section{The massless model}
\indent

The interaction of massless right and left chiral 
fermions ($\psi$ and $\lambda$) of charges
$e$ and $e'$ with a U(1) gauge field ($A_\mu$) is
described by the Lagrangian density 
\footnote{In our notation \cite{aar}, $\gamma^0=\big(\matrix{&0 &1\cr
&1 &0\cr}\big)$,  $\gamma^1=\big(\matrix{&0 &-1\cr
&1 &0\cr}\big)$,  $\gamma^5=\big(\matrix{&1 &0\cr
&0 &-1\cr}\big)$, $\gamma^\mu\gamma^5 =\epsilon^{\mu\nu}\gamma_\nu$,
$\epsilon_{01}=+1$ and $\tilde p_\mu=\epsilon_{\mu\nu}p^\nu$. }
\begin{eqnarray}
{\cal L}&=&-{1\over 4} F_{\mu\nu} F^{\mu\nu}
+\sum_{j=1}^n\bar\psi_j\gamma^\mu(i\partial_\mu+e A_\mu {1+\gamma_5\over
2})\psi_j \nonumber
\\
& & +\sum_{l=1}^q \bar\lambda_l\gamma^\mu(i\partial_\mu+
e'A_\mu{1-\gamma_5\over
2})\lambda_l,\label{lagrangian}
\end{eqnarray}
where the space-time indices are $\mu,\nu=\{0, 1\}$.
The coupling of the gauge field to the non-conserved currents,
\be
J^+= {e\over 2}\sum_j\bar\psi_j\gamma^+\psi_j = e \psi_-^+\psi_- 
\label{jplus}
\ee
and
\be
J^-= {e'\over 2}\sum_l\bar\lambda_l\gamma^-\lambda_l = e' \lambda_+^+
\lambda_+\label{jminus}
\ee
renders this model anomalous. The breakdown of the gauge 
symmetry is made manifest in the anomaly
equation which expresses the non-vanishing of
the total divergence of the chiral currents, {\it i.e.},
\be
\partial^\mu J_\mu = {1\over 4\pi} (n e^2 - q e'^2)\epsilon^{\mu\nu} 
F_{\mu\nu}.\label{anomaly}
\ee

A well-known method of dealing with an anomalous model 
is to introduce extra degrees of
freedom \cite{ht} into the theory. In this procedure, 
the chiral theory is replaced 
by a gauge-invariant theory
which incorporates the additional modes.
The original anomalous theory can be
recovered by a suitable choice of gauge \cite{aar}.
However, it is believed that a reliable physical theory 
cannot contain gauge anomalies. 
The gauge symmetries can be restored by imposing 
the anomaly cancellation condition on the
theory. Specifically, such a cancellation is achieved
by requiring that the 
coefficient of the anomaly equation
vanishes, {\it i.e.},
\be
ne^2=q{e'}^2\Rightarrow({e\over e'})^2={q\over n}. \label{matching}
\ee
We shall study this anomaly-free
quantum model and solve it in terms
of elementary quantum scalar fields.

\section{Quantum solution of the massless theory}

\indent

In order to obtain the complete operator solution to the model 
(\ref{lagrangian}), we need
to solve the equations of motion and to insure that the
physical Hilbert space is consistent with Maxwell equations.

The Dirac equations of motion, obtained from the Lagrangian 
(\ref{lagrangian}) is 
\footnote{Note that
the Dirac equations are similar to those obtained for 
flavoured Schwinger model.}
\be
i\partial\eta +{\tilde e}\, /\hskip -2.5mm A{1+\epsilon\gamma_5\over 2}\eta=0,
\label{dirac}
\ee
where $\eta$ is $\psi_i$ or $\lambda_l$, $\epsilon=\pm 1$ and $\tilde e=e, e'$.
The most general expression for the gauge field is
\be
A_\mu=-(\gamma{\tilde\partial}_\mu\Sigma+\delta\partial_\mu\eta),
\label{gaugefield}
\ee
where $\Sigma$ and $\eta$ are any scalar fields.
Consequently, the field strength takes the form,
\be
F_{\mu\nu}=\gamma\epsilon_{\mu\nu}\Box\Sigma.\label{fmunu}
\ee
Similarly, the fermions can be expressed in terms of bosonic fields
$\varphi_f$, $\zeta_f$, $\eta$ and $\Sigma$ as follows 
\footnote{The fields $\Sigma$ and $\eta$
do not appear in the expressions for 
$\psi_+$ or $\lambda_-$ which
are the free chiral components.}:
\begin{eqnarray}
\psi_{+,f}&=&({\mu\over 2\pi})^{1/2} K_f \exp\bigg[{i\over
4}\pi+i\sqrt\pi\big(-\varphi_f
+
\int_{x^1} dy^1 \dot\varphi_f(x^0,y^1)\big)
\bigg],\nonumber
\\
& & \label{psiplus}
\\
\psi_{-,f}&=&\big({\mu\over 2\pi}\big)^{1/2} K_f \exp\bigg[-{i\over
4}\pi+i\sqrt\pi\big(\varphi_f
+ \int_{x^1} dy^1 \dot\varphi_f(x^0,y^1)\big)\bigg]\nonumber
\\
&\times &
\exp\bigg[i\big(e\gamma\Sigma+e\delta\eta\big)\bigg],\label{psiminus}
\\
\lambda_{+,f^\prime}&=&({-\mu\over 2\pi})^{1/2} K_f \exp\bigg[{i\over
4}\pi+i\sqrt\pi\big(-\zeta_f
+
\int_{x^1} dy^1 \dot\zeta_{f^\prime}
(x^0,y^1)\big)\bigg]\nonumber
\\
&\times &
\exp\bigg[-i\big(e'\gamma^\prime\Sigma+e'\delta^\prime\eta^\prime\big)\bigg],
\label{lambdaplus}\\
\lambda_{-,f^\prime}&=&({\mu\over 2\pi})^{1/2} K_f \exp\bigg[{-i\over
4}\pi+i\sqrt\pi\big(-\zeta_{f^\prime}+\int_{x^1} dy^1 
\dot\zeta_{f'}(x^0,y^1)\big)\bigg],\nonumber
\\
& &\label{lambdaminus}
\end{eqnarray}
where the Klein-factors $K_f$ insure the correctness of the
commutation relations \cite{aar}.
The gauge and the fermion fields
satisfy the Dirac equations provided $\gamma=\gamma^\prime$,
$\delta=\delta^\prime$ and the fields 
$\zeta_f$, $\varphi_f$ and $\eta$ are
massless. 

In addition to the Dirac equations, Maxwell equations,
\be
\partial_\mu F^{\mu\nu}+e\sum_j \bar\psi\gamma^\nu{1+\gamma_5\over 2}\psi
+ e'\sum_l\bar\lambda\gamma^\nu{1-\gamma_5\over 2}\lambda=0,\label{maxwell}
\ee
should also be satisfied \footnote{In fact, Maxwell 
equations can only be satisfied 
on the physical Hilbert space.}. Using eqs. 
(\ref{psiplus}-\ref{lambdaminus}), and after some manipulations
\cite{aar}, we obtain the currents,  
\begin{eqnarray}
J_-^\psi&=&\bar\psi\gamma_-\psi=-{1\over\sqrt\pi}\partial_-\varphi_f
-{e\over\pi}\delta\partial_-\eta-{e\over\pi}\gamma\partial_-\Sigma ,
\label{jpsiminus}
\\
J_+^\psi&=&\bar\psi_f\gamma_+\psi_f={1\over\sqrt\pi}
\partial_+\varphi_f,\label{jpsiplus}
\\
J_-^\lambda&=&-{1\over\sqrt\pi}\partial_-\zeta_f,\label{jlambdaminus}
\\
J_+^\lambda&=&{1\over\sqrt\pi}\partial_+\zeta_f+
{e'\over\pi}\delta\partial_+\eta +{e'\over\pi}\gamma\partial_+\Sigma .
\label{jlambdaplus}
\end{eqnarray}
On inserting these currents into the Maxwell 
equations (\ref{maxwell}), we obtain for $\nu=-$, 
\be
\gamma\partial_+\Box\Sigma+{e'\over\sqrt\pi}\partial_+\sum_f\zeta_f
+{e'\over\pi}q^2\delta\partial_+\eta+
e{q^\prime}^2{\gamma\over\pi}\partial_+\Sigma=0,\label{numinus}
\ee
and for $\nu=+$,
\be
-\gamma\partial_-\Box\Sigma-{e\over\sqrt\pi}\partial_-\sum_f\varphi_f
-{1\over\pi}\delta e^2 n\partial_-\eta-{\gamma\over\pi}n
e^2\partial_-\Sigma=0.\label{nuplus}
\ee
The massless terms ({\it i.e.} the terms containing
massless fields) in the above two equations have to vanish 
separately. This is only possible
if the space of states is chosen in such a way that
the longitudinal currents
\be
L_+=\partial_+\big[{e'\over\sqrt\pi}\sum_f\zeta_f
+{qe'^2\over\pi}\delta\eta\big]\label{lplus}
\ee
and
\be
L_-=\partial_-\big[{e\over\sqrt\pi}\sum_f\varphi_f
+{qe^2\over\pi}\delta\eta\big]\label{lminus}
\ee
vanish weakly, {\it i.e.},
the physical subspace is defined to
be that generated by states which obey
\be
\langle{\rm phys}\vert L_\mu\vert{\rm phys}\rangle=0.\label{guptableuler}
\ee
In other words, the Gupta-Bleuler (or BRST) condition must be satisfied. 

In contrast to the fields $\eta$, $\zeta_f$ and $\varphi_f$, which are
massless, the bosonic field $\Sigma$
obeys a massive equation. For this field,
the contradictory equations of motion,
\be
(\gamma\Box+{\gamma\over\pi} q e'^2)\Sigma=0\label{contra1}
\ee
and
\be
(\gamma\Box +{\gamma\over\pi} n e^2)\Sigma=0,\label{contra2}
\ee
only coincide if the anomaly-cancellation condition
(\ref{matching}) is imposed on the charges $e$ and $e'$. 
Consequently, the mass of the $\Sigma$ field is \footnote{
This corresponds to the dynamical Higgs mechanism.}
\be
m^2_\Sigma={ne^2\over\pi}={qe'^2\over\pi}.\label{masssigma}
\ee

The parameters $\gamma$ and $\delta$ can now be computed.
We require the vanishing of the norm of $L_\mu$
and impose canonical quantisation condition 
on $\eta$, taking care of the negative metric
character of the commutator
\footnote{The
fields $\zeta_f$ and $\varphi_f$ are also canonically quantised, but with 
positive metric.}.
We then obtain 
\be
{ne^2\over\pi}\big[4-{ne^2\delta^2\over\pi}\big]=0
\Rightarrow\delta={1\over e}\sqrt{\pi\over n}={1\over e'}\sqrt{\pi\over
q}\label{delta}
\ee
Similarly, by imposing the canonical
commutation relation on the gauge field $A_1$ \footnote{The gauge field 
is $A_1=\gamma\partial_0\Sigma$
and the momentum conjugate to it is
$\pi_1=F^{01}=\gamma\Box\Sigma=-\gamma{ne\over\pi}\Sigma$.} and 
on the scalar field $\Sigma$, we obtain
\be
\gamma={1\over e}\sqrt{\pi\over n}={1\over e'}\sqrt{\pi\over q}.\label{gamma}
\ee

Next, we construct the physical operators.  
We define $\varphi={1\over\sqrt
n}\sum_f\varphi_f$ and
$\zeta={1\over\sqrt q}\sum_{f\p}\zeta_{f\p}$. The
longitudinal currents (\ref{lplus}, \ref{lminus}) simplify to
\be
L_+=e\p\sqrt{q\over\pi}\partial_+(\zeta+\eta)\label{lplus2}
\ee
and
\be
L_-=e\sqrt{n\over\pi}\partial_-(\varphi+\eta).\label{lminus2}
\ee
The negative metric part of the theory can now be isolated
by re-expressing the physical operators in terms of the combined fields
$\varphi+\eta$ and $\zeta+\eta$.
The field $\psi_-$ depends on $\varphi$ and $\eta$ and the field
$\lambda_+$ 
on $\zeta$ and $\eta$. However, physical quantities can only depend
on the combinations $\varphi +\eta$ or $\zeta+\eta$, due to the physical
state condition (\ref{guptableuler}). Also, the fields with opposite 
chiralities $\psi_+$ and $\lambda_-$ should remain invariant. Thus we need
a chiral gauge transformation. Taking these considerations
into account, we define the following transformations \footnote{The partition 
function remains invariant as a result
of the fine tuning of the charges $e$ and $e'$.}: 
\begin{eqnarray}
\psi_-\rightarrow e^{i\sqrt{\pi\over n}\int_{x^1}\dot\eta}\psi_- \quad , 
\quad\psi_+\rightarrow\psi_+ ,\nonumber\\
\lambda_+\rightarrow e^{-i\sqrt{\pi\over q}
\int_{x^1}\dot\eta}\lambda_+\quad  , 
\quad\lambda_-\rightarrow\lambda_- ,\nonumber\\
A_-\rightarrow A_--{1\over e}\sqrt{\pi\over n}\partial_-\eta ,\nonumber\\
A_+\rightarrow A_++{1\over e}\sqrt{\pi\over n}\partial_+\eta.
\label{chiralgauge}
\end{eqnarray}
The physical fields are now given by,
\begin{eqnarray}
{\psi_-}_f&=&({\mu\over 2\pi})^{1/2} K_f e^{i\sqrt{\pi\over n}\Sigma}\nonumber
\\
&\times & \exp {i\sqrt\pi\big[\varphi_f-{1\over\sqrt n}\varphi +\int 
(\dot\varphi_f -{1\over\sqrt n}\dot\varphi)\big]}\nonumber
\\
&\times&\exp{i\sqrt{\pi\over n}\big[\eta+\varphi+\int(\dot\eta+
\dot\varphi)}\big]\label{psiminusphys}
\end{eqnarray}
and
\begin{eqnarray}
{\lambda_+}_{f^\prime}&=&({\mu\over 2\pi})^{1/2} K_{f^\prime} 
e^{-i\sqrt{\pi\over
q}\Sigma}\nonumber
\\
&\times&\exp{i\sqrt\pi\big[-\zeta_{f^\prime}+{1\over\sqrt q}\zeta + \int
(\dot\zeta_{f^\prime}-{1\over\sqrt q}\zeta)\big]}\nonumber
\\
&\times& \exp{-i\sqrt{\pi\over q}
\big[\eta+\zeta-\int(\dot\eta+\dot\zeta)\big]},\label{psiplusphys}
\end{eqnarray}
which depend on the required combination of fields.
The last exponentials appearing in the expressions for the
physical fields, {\it i.e.},
\be
\sigma(x)=\exp{i\sqrt{\pi\over
n}\big[\eta+\varphi+\int(\dot\eta+\dot\varphi)\big]}=
e^{i\theta_1}\label{theta1}
\ee
and
\be
\rho(x)=\exp{-i\sqrt{\pi\over
q}\big[\eta+\zeta-\int(\dot\eta+\dot\zeta)\big]}=
e^{i\theta_2},\label{theta2}
\ee
act as constant operators on the physical Hilbert space.\footnote{
This can be shown by using the two-point-functions relations,
\begin{eqnarray}
\langle\varphi(x)\varphi(y)\rangle&=&
\langle\zeta(x)\zeta(y)\rangle=-\langle
\eta(x)\eta (y)\rangle\,.
\nonumber
\end{eqnarray}
}
This gives rise to the angles
$\theta_1$ and $\theta_2$. Thus, the vacuum has a 
structure similar to that of 
the Schwinger model--it is degenerate and 
is labelled by the angles $\theta_1$ and $\theta_2$.

For general values of $n$ and $q$, the solutions are
similar to those of the flavoured Schwinger model. 
However, for rational values of $\sqrt{n/q}$ a special situation arises.
One can construct condensates in the following way:
\be
\eta_{f_1,\cdots f_{\sqrt {n\p}}f_1\p,\cdots f_{\sqrt {q\p}}\p}=
\psi_{f_1}\cdots\psi_{f_{\sqrt n\p}}\,
\bar\lambda_{f_1\p}\cdots\bar\lambda_{f_{\sqrt{q\p}}\p},\label{condensate}
\ee
where $n\p$ and $q\p$ are such that $\sqrt{n\p}$ and $\sqrt{q\p}$
are integers and obey $\sqrt{n\p/q\p}=\sqrt{n/q}$. It is also possible to 
define the duals
\be
\tilde \eta_{f_1,\cdots f_{n\p -\sqrt {n\p}}f_1\p,\cdots 
f_{q\p -\sqrt {q\p}}\p}=
\psi_{f_1}\cdots\psi_{f_{n\p -\sqrt n\p}}\,
\bar\lambda_{f_1\p}\cdots\bar\lambda_{f_{q\p -\sqrt{q\p}}\p},
\label{dualcondensate}
\ee

The $\eta$--$\tilde\eta$-operators have
non-vanishing two-point functions \break
$\langle\eta_{\{f\}\{f\p\}}(x)
\tilde\eta_{\{\tilde f\}\{\tilde {f\p}\}}(y)\rangle$ which
break the fermion-number conservation law\cite{nn}
\footnote{Such two-point functions can be computed 
by known techniques of
two-dimensional quantum field theory \cite{aar}.}. 
Let us take the example of the $SU(4)_R\times U(1)_L$
theory, which contains particles of the type
\be
\eta(x)_{ij}=\psi_i\psi_j\overline\lambda(x)\label{2condensate}
\ee
and where the dual coincides with the original condensate, after a
permutation of indices.  We are interested in computing the correlators
for long distances. In that case, we can drop the contribution of the
massive field, and we are left with
\begin{eqnarray}
\lefteqn{\langle\eta_{12}(x)\eta_{34}(0)\rangle\approx}\nonumber\\
&\approx&\big\langle 
e^{-{i\over 2}\sqrt\pi(\varphi_1^{(+)}+
\varphi_2^{(+)}-\varphi_3^{(+)}-\varphi_4^{(+)})(x)}
e^{-{i\over 2}\sqrt\pi(\varphi_1^{(+)}+
\varphi_2^{(+)}-\varphi_3^{(+)}-\varphi_4^{(+)})(0)}\big\rangle,\nonumber \\
&=&e^{-{1\over 4}
4{\rm ln}(-x)} = \lbrack -x^-\rbrack^{-1}\quad .\label{condenscorrelator}
\end{eqnarray}
where the label-$+$ means the combination $\varphi^{(+)}=\varphi
+\int_{x^1}dy^1\dot \varphi(x^0,y^1)$. Notice the use of the infrared
regularization of the exponential of the massless field in two dimensions,
crucial in the present computation.
In general, such two-point correlators are expected to vanish. Here they do
not, in view of instanton effect as already announced in \cite{nn}.
These excitations can be created 
in a world where $\sqrt{n/q}$ takes on rational values.
Moreover, these condensates only have a $-1$ power decay (that is as a 
fermion) for very particular models, such as the example $SU(4)\times U(1)$
above, discussed in \cite{nn}. If we take the example of a model
$SU(9)\times U(1)$, we have for the corresponding correlator the expression
\be
\langle\eta_{123}(x)\tilde\eta_{456789}(0)\rangle\approx 
\lbrack -x^-\rbrack^{-1}\quad . \label{condenscorrelator9-1}
\ee
For a general symmetry $SU(n^2)\times SU(q^2)$, with both $n$ and $q$
not unit, we obtain
\be 
\langle\eta_{ff\p}(x)\tilde\eta_{\tilde f\tilde f\p}(0)\rangle\approx 
\lbrack -x^-\rbrack^{-1}\lbrack -x^+\rbrack^{-1}\quad .
\label{generalcorrecondens}
\ee
Analogously, for a symmetry $SU(n^2)\times U(1)$ a correlator
\be
\langle \eta (x_1)\cdots\eta (x_n)\rangle\approx \prod_{i<j} 
\lbrack x_i^- -x_j^- \rbrack^{-1} 
\ee
where all indices are different does not vanish either.

\section{The massive theory}
\indent

Unlike the massless theory,
the massive theory is not exactly solvable. 
However, to obtain a bosonised massive lagrangian, 
the mass term of the original lagrangian can be treated 
perturbatively. Mass terms of the bosonised lagrangian 
are constructed by using the fields 
appearing in the massless bosonised theory. In addition, one should
allow for possible renormalisation constants. This procedure is 
known as the principle of form invariance \cite{cjs}.

The massive theory is described by the lagrangian
\be
{\cal L}_{\rm m}={\cal L}+{\cal L}_{{\rm mass}}\,,
\ee
where ${\cal L}$ is given by (\ref{lagrangian}) and 
\be
{\cal L}_{{\rm mass}} =
m\sum_f\psi_{+f}^\dagger\psi_{-f}+m'\sum_{f'}
\lambda^\dagger_{+f}\lambda_{-f} + c.c.\, .\label{lzeromass}
\ee
The combinations $\psi_+^\dagger\psi_-$ or
$\lambda_+^\dagger\lambda_-$ in the mass terms 
are determined by the
charge superselection rules. The massive lagrangian can 
be bosonised by using the formulae 
(\ref{psiplus}-\ref{lambdaminus}), 
obtained in the massless case. We find,
\begin{eqnarray}
{\cal L}^{(0)}_{{\rm mass}}&=&m\sum_f{\rm cos}\big[2\sqrt\pi
\varphi_f+\sqrt{\pi\over n}(\eta+\Sigma)\big]\nonumber\\
&+&m^\prime\Sigma_{f^\prime}{\rm cos}\big[2
\sqrt\pi\zeta_f +\sqrt{\pi\over
q}(\eta+\Sigma)\big].\label{lzeromassbos}
\end{eqnarray}
The bosonised massive lagrangian obtained 
from a simple extension of the
massless lagrangian is, however, incomplete.
The equations of motion are
inconsistent and the physical operators 
fail to satisfy the BRST conditions.

Nevertheless, expression (\ref{lzeromassbos}) gives us 
an insight into the possible form of
the correct lagrangian. It shows that
all the fields in the massive 
theory are interacting. We also use
the principle of form invariance, discussed above,  to
propose the ansatz
\begin{eqnarray}
{\cal L}_{{\rm mass}}&=& m\sum_f{\rm
cos}\big[\alpha\varphi_f+\beta\eta+\gamma\Sigma+
\sigma\Sigma+\sigma\varphi+\epsilon\zeta\big]\nonumber
\\
&+& m^\prime\sum_{f^\prime}{\rm
cos}\big[\alpha^\prime\varphi_{f^\prime}+\beta^\prime\eta+\gamma^\prime\Sigma+
\sigma^\prime\Sigma+\sigma^\prime\varphi+\epsilon^\prime\zeta\big].
\label{lmass}
\end{eqnarray}
The equations of motion obtained from the total lagrangian which includes the
above mass terms are,
\begin{eqnarray}
(\Box+\mu^2)\Sigma&+&m\gamma\sum_f{\rm
sin}\big[\alpha\varphi_f+\beta\eta+\gamma\Sigma+
\sigma\varphi+\epsilon\zeta\big]\nonumber
\\
&+& m^\prime\gamma^\prime\sum_{f^\prime}{\rm
sin}\big[\alpha^\prime\zeta_{f^\prime}+\beta^\prime\eta+\gamma^\prime\Sigma+
\sigma^\prime\varphi+\epsilon^\prime\zeta\big]=0,\label{sigmamasseq}
\\
\Box\varphi_f &+&
m\alpha\sum_f{\rm
sin}\big[\alpha\varphi_f+\beta\eta+\gamma\Sigma+
\sigma\varphi+\epsilon\zeta\big]\nonumber
\\
&+&
m{\sigma\over\sqrt n}\sum_{\tilde f}{\rm
sin}\big[\alpha\varphi_{\tilde f}+\beta\eta+\gamma\Sigma+
+\sigma\varphi+\epsilon\zeta\big]\nonumber
\\
&+& m^\prime\alpha^\prime\sum_{{\tilde f}^\prime}{\rm
sin}\big[\alpha^\prime\zeta_{{\tilde f}^\prime}+\beta^\prime\eta
+\gamma^\prime\Sigma+
\sigma^\prime\zeta+\epsilon^\prime\varphi\big]=0,
\label{varphimasseq}
\\
\Box\zeta_f &+&
m\p\alpha\p\sum_f{\rm
sin}\big[\alpha\p\varphi_f+\beta\p\eta+\gamma\p\Sigma+
\sigma\p\zeta+\epsilon\p\varphi\big]\nonumber
\\
&+&
m\p{\sigma\p\over\sqrt q}\sum_{{\tilde f}\p}{\rm
sin}\big[\alpha\p\zeta_{{\tilde f}\p}+\beta\p\eta+\gamma\p\Sigma+
\sigma\p\zeta+\epsilon\p\varphi\big]\nonumber
\\
&+& m\alpha\sum_{\tilde f}{\rm
sin}\big[\alpha\varphi_{{\tilde f}}+\beta\eta
+\gamma\Sigma+
\sigma\varphi+\epsilon\zeta\big]=0,\label{zetamasseq}
\\
-\Box\eta&+&m\beta\sum_f{\rm
sin}\big[\alpha\varphi_f+\beta\eta+\gamma\Sigma+
\sigma\varphi+\epsilon\zeta\big]\nonumber
\\
&+& m^\prime\beta^\prime\sum_{f^\prime}{\rm
sin}\big[\alpha^\prime\zeta_{f^\prime}+\beta^\prime\eta+\gamma^\prime\Sigma+
\sigma^\prime\varphi+\epsilon^\prime\zeta\big]=0 .
\label{etamasseq}
\end{eqnarray}
The equations of motion of the fields
which are relevant for the BRST conditions
follow from these;
\begin{eqnarray}
\Box(\varphi+\eta)&=&-m({\alpha\over\sqrt n}+\sigma-\beta)\sum_f{\rm
sin}\big[\alpha\varphi_f+\beta\eta+\gamma\Sigma+
\sigma\varphi+\epsilon\zeta\big]\nonumber
\\
&-& m^\prime(\epsilon\p-\beta^\prime)\sum_{f^\prime}{\rm
sin}\big[\alpha^\prime\zeta_{f^\prime}+\beta^\prime\eta+\gamma^\prime\Sigma+
\sigma^\prime\varphi+\epsilon^\prime\zeta\big],\label{varphipluseta}
\\
\Box(\zeta+\eta)&=&-m\p({\alpha\p\over\sqrt q}+\sigma\p-\beta\p)\sum_{f\p}
{\rm sin}\big[\alpha\p\zeta_f+\beta\p\eta+\gamma\p\Sigma+
\sigma\p\varphi+\epsilon\p\zeta\big]\nonumber
\\
&-& m(\epsilon-\beta)\sum_f{\rm
sin}\big[\alpha\varphi_f+\beta\eta+\gamma\Sigma+
\sigma\varphi+\epsilon\zeta\big].\label{zetapluseta}
\end{eqnarray}
The above combined operators are required, by the
Gupta-Bleuler condition,
to be free massless fields (see eqs. (\ref{lplus2}) and (\ref{lminus2})). 
Therefore, we obtain
\be
\beta={\alpha\over\sqrt n}+\sigma=\epsilon\label{beta}
\ee
and
\be
\beta\p={\alpha\p\over\sqrt q}+\sigma\p=\epsilon\p.\label{betaprime}
\ee

The dimensions of the mass operators,
\begin{eqnarray}
{\rm dim}\,(\bar\psi\psi)&=&{\alpha^2+2\alpha\sigma/\sqrt
n+\sigma^2+\gamma^2+\epsilon^2-\beta^2\over 4\pi},\nonumber
\\
&=&
{\alpha^2+\sigma^2+\gamma^2+2\alpha\sigma/\sqrt n\over 4\pi},\label{dim2psi}
\\
{\rm dim}\,(\bar\lambda\lambda)&=&{{\alpha\p}^2+\sigma^{\p 2}+\gamma^{\p
2}+2\alpha\p\sigma\p/\sqrt n\over 4\pi},\label{dim2lambda}
\end{eqnarray}
determine the range of the allowed values of the parameters $\gamma$
and $\gamma\p$. 
If, as in the massive Schwinger model, we require the absence of 
Thirring interactions and fix the parameter $\alpha$ to be
$2\sqrt\pi$, the mass operator will have 
dimension greater than one \cite{aar}. On evaluating the 
operator-product expansion of the canonical fermion fields,
we also find that $\sigma\p=\sigma=0$.
Under such conditions, we obtain 
\begin{eqnarray}
{\rm dim}\,(\bar\psi\psi)&=&{\alpha^2+\gamma^2\over 4\pi}\rightarrow
1+{\gamma^2\over 4\pi}\,,\label{dim2psif}
\\
{\rm dim}\,(\bar\lambda\lambda)&\rightarrow& 1+{{\gamma\p}^2\over 4\pi}\,.
\label{dim2lambdaf}
\end{eqnarray}
The mass perturbation expansion is only 
well-defined if the dimensions of these mass
operators do not exceed 2. This is required for renormalizability. Therefore,
we conclude that, $\gamma\leq 2\sqrt\pi$ and $\gamma\p\leq 2\sqrt\pi$.

\section{Conclusion and Comments}
\indent 

We have constructed the full operator solutions
of chiral gauge theories and obtained the anomaly-cancellation conditions
by requiring the consistency of these solutions. The solutions together 
with the structure of the 
$\theta$-vacuum are similar to their counterparts in the Schwinger model.
It would be interesting to see if these theories also have phase structures
similar to the screening and confining phases of the Schwinger model.

For symmetry groups $SU(n)_R\times SU(q)_L$ with $\sqrt{n/q}$ 
a rational number, we have constructed condensates. For long distances these
condensates behave like fermions.
We have shown that they violate 
fermion-number conservation law. It remains to be seen if this 
phenomenon can be generally and satisfactorily explained 
in terms of instantons, as discussed in the case of symmetry 
$SU(4)\times U(1)$. In the bosonised theory, the fermion number is a 
topological number and the
instantons can interplay between different vacua. Thus, the
non-conservation of the fermion number can be traced back to
instanton effects.
The generalization of these results to a nonabelian gauge group would
be welcome,\cite{ea} but the nonabelian bosonisation formulas
are more involved, and do not permit a straightforward solution.

We have formulated a bosonised massive lagrangian by using the principle 
of form invariance, the BRST condition and mass perturbation techniques. 
However, it remains a challenging open problem to find a
bosonisation formula which represents massive fermions as 
operator functions of bosonic variables.

\subsubsection*{Acknowledgements} 
I wish to thank R. Mohayaee, J. Narayanan, H. Neuberger
and K.D. Rothe for discussions and suggestions, which considerably
improved this paper.



\begin{thebibliography}{99}
\frenchspacing

\bibitem{sch}J. Schwinger, {\it Phys. Rev.} {\bf 128} (1962) 2425.\\
\bibitem{ls}J. Lowenstein and J.A. Swieca, {\it Annals of Phys.} {\bf 68} 
(1971)
172.\\
\bibitem{cks}A. Casher, J. Kogut and L. Susskind, {\it Phys. Rev } {\bf D10}
(1974) 732.\\
\bibitem{aar}E. Abdalla, M. C. B. Abdalla and K. D. Rothe, 
``Non-perturbative methods in two dimensional quantum field theory'', 
World Scientific, 1991.\\
\bibitem{ht}K. Harada and I. Tsutsui, {\it Phys. Lett.} {\bf B183} (1987) 
311.\\
\bibitem{nn}J. Narayanan and H. Neuberg, RU-96-84 Preprint, hep-lat/9609031.\\
\bibitem{ea}E. Abdalla and M. C. B. Abdalla, {\it Phys. Rep.} {\bf 265}
 (1996) 253.\\

\bibitem{cjs} S. Coleman, R. Jackiw and L. Susskind, {\it Ann. Phys.}
 {\bf 13} (1975) 267.
 
\end{thebibliography}
\end{document}